\documentclass[prb,amsmath,amssymb,showpacs,twocolumn,letterpaper]{revtex4}
\usepackage{graphicx}
\usepackage{CJK}
\begin{document}
\begin{CJK*}{UTF8}{gbsn}
\title{Thermal fluctuations and flux-tunable barrier in proximity Josephson junctions}

\author{Jian Wei (危健)}
\email{weijian6791@pku.edu.cn}
\altaffiliation[Present address: ]{School of Physics, Peking University, P.R. China}
\affiliation{Department of Physics and Astronomy, Northwestern University, Evanston, IL 60208, USA}
\author{P. Cadden-Zimansky}
\altaffiliation[Present address: ]{Department of Physics, Columbia University, New York, NY 10027, USA}
\affiliation{Department of Physics and Astronomy, Northwestern University, Evanston, IL 60208, USA}
\author{P. Virtanen} 
\altaffiliation[Present address: ]{Institute for Theoretical Physics and Astrophysics,
University of W\"{u}rzburg, D-97074 W\"{u}rzburg, Germany.}
\affiliation{Low Temperature Laboratory, Helsinki University of Technology, Helsinki, Finland}
\author{V. Chandrasekhar}
\email{v-chandrasekhar@northwestern.edu}
\affiliation{Department of Physics and Astronomy, Northwestern University, Evanston, IL 60208, USA}

\date{\today}


\begin{abstract}
The effect of thermal fluctuations in Josephson junctions is usually analysed using the Ambegaokar-Halperin (AH) theory in the context of thermal activation. ``Enhanced'' fluctuations, demonstrated by broadening of current-voltage characteristics, have previously been found for proximity Josephson junctions. Here we report measurements of micron-scale normal metal loops contacted with thin superconducting electrodes, where the unconventional loop geometry enables tuning of the junction barrier with applied flux; for some geometries, the barrier can be effectively eliminated. Stronger fluctuations  are observed when the flux threading the normal metal loop is near an odd half-integer flux quantum, and for devices with thinner superconducting electrodes. These findings suggest that the activation barrier, which is the Josephson coupling energy of the proximity junction, is different from that of conventional Josephson junctions. Simple one dimensional quasiclassical theory can predict the interference effect due to the loop structure, but the exact magnitude of the coupling energy cannot be computed without taking into account the details of the sample dimensions.  In this way, the physics of this system is similar to the phase slipping process in thin superconducting wires. Besides shedding light on thermal fluctuations in proximity junctions, the findings here also demonstrate a new type of superconducting interference device with two normal branches sharing the same SN interface on both sides of the device, which has technical advantages for making symmetrical interference devices.
\end{abstract}

\maketitle
\end{CJK*}

Understanding the effect of thermal fluctuations in nanoscale superconducting devices is important for applications.\cite{Giazotto2006rmp} One particular type of nanoscale superconducting device is a proximity junction,\cite{Giazotto2008,Wei2008a} consisting of a normal metal wire contacted by two superconducting wires as electrodes. The difference between such a device and a conventional Josephson junction made of bulk superconducting electrodes separated by a thin insulator is twofold: 1) Instead of tunnelling through the insulating barrier, charge carriers diffuse through the normal metal barrier in proximity junctions, leading to a stronger coupling between the two superconducting electrodes for proximity junctions;  2) the thin superconducting electrodes of these nanoscale junctions are different from bulk superconductors since their dimensions are now comparable to the superconducting coherence length $\xi_{s}$. 

These differences may explain the previously observed strong fluctuations in nanoscale proximity junctions,\cite{Hoss2000,Dubos2001b,Lhotel2007} which manifest themselves as ``enhanced'' broadening of the current-voltage characteristics (CVC) compared to the ``intrinsic'' broadening of CVC due to thermal activation of a resistively shunted junction (RSJ), as described by Ambegaokar and Halperin (AH).\cite{Ambegaokar1969,Ivanchenko1968} This enhancement can be characterized by an effective noise temperature $T_{N}$, which is higher than the bath temperature $T_{b}$.\cite{Akazaki2005}  

In this paper, we report measurements of a special type of proximity junction with a loop structure embedded into the junction itself. By threading a magnetic flux through the loop, we find $T_{N}$ is maximum when the flux through the loop is an odd half-integer of the superconducting flux quantum $\Phi_0=h/2e$, for fixed $T_{b}$.  The ratio between $T_{N}$ and $T_{b}$ depends on the geometries of the particular sample, but does not depend on the bath temperature $T_{b}$. The flux dependence of $T_{N}$ is better understood if we consider a phase slipping process similar to that in thin superconducting wires, which again suggests that for a nanoscale proximity junction, the activation energy is different from the standard Josephson coupling energy.  Such devices exhibit many of the properties of dc SQUIDs, but with the advantage that the devices can be designed to be almost perfectly symmetric, as the two normal branches share the same SN interface on both sides of the device, hence allow unprecedented tunability of the system by means of an external magnetic flux.

The remainder of this paper is organized as follows: In Section I, a general introduction for thermal activation in superconducting wires and Josephson junctions is presented, followed by a discussion of how quasiclassical theory can be used to extend the thermal activation model to proximity junctions. Then the experimental results of magnetoresistance (Section II), the current voltage characteristics (CVC) (Section III), and temperature dependence of resistance (Section IV) are discussed. Finally we summarize the findings and propose further theoretical study beyond one dimensional quasiclassical theory. 

\section{Theoretical background}

The dynamics of a particle trapped in a shallow potential well is a fundamental problem that has wide applicability to a number of areas in statistical physics.  At finite temperatures, due to coupling to a thermal bath, the particle may escape from the potential well through the process of thermal activation over the barrier represented by the edge of the potential.  

For superconductors, two specific phenomena have been explored extensively in terms of the physics of thermal activation.  The first phenomenon is the generation of phase-slips in a thin superconducting wire with cross-section size comparable to the superconducting coherence length $\xi_S$.\cite{Little1967,Langer1967}  In this case,  thermally activated phase slips (TAPS) lead to the appearance of a resistance tail of the thin wire at temperatures below the nominal transition temperature $T_c$.  The voltage $V$ generated by TAPS is related to the time evolution of the macroscopic superconducting phase $\varphi$  through the Josephson relation $2eV/\hbar = d\varphi/dt$, which on average is determined by the number of TAPS per unit time, each TAPS corresponding to a change of $2 \pi$.  At the instant in space and time where such a phase slip event occurs, the superconducting order parameter vanishes, which costs an energy $\Delta F$, the barrier over which the system must be thermally activated. In the earlier pioneering theory by Little, \cite{Little1967}  the relevant energy barrier is the condensation energy in a small coherent volume determined by the dimensions of the superconducting wire, and the temperature dependence of resistance $R(T)$ is described by an Arrhenius type equation. Later developments gave a more accurate description of $\Delta F$ and the attempt frequency for activation over the barrier, often referred to as the LAMH theory.\cite{Langer1967,McCumber1970} 

The second phenomenon is the onset of finite voltage in a Josephson junction.  A resistively shunted junction (RSJ) can be modelled as a particle in a one-dimensional washboard potential in a viscous medium, where the distance coordinate corresponds to the phase difference $\varphi$ across the junction, as shown by Ambegaokar and Halperin (AH). \cite{Ambegaokar1969,Ivanchenko1968}  With no current through the junction, the system sits in a local minimum of the potential.  Application of a current through the junction corresponds to tilting the washboard, shifting the position of the local minima.  At some value of current less than the nominal critical current $I_c$, the washboard potential is tilted sufficiently for the system to be thermally activated over the barrier between two adjacent potential minima.  Once this occurs, the system will continue to roll down the washboard potential, corresponding to a continuous time evolution of $\varphi$, and hence a finite voltage will appear across the junction according to the Josephson relation.  Here the energy barrier (the height of the washboard potential) being thermally activated over is the Josephson coupling energy $E_J = (\hbar/2e) I_c$, where $I_c$ is the critical current.

A proximity junction made by a normal metal wire between two superconductors exhibits properties different from a conventional tunnelling junction. Proximity SNS junctions have also been investigated for a long time.\cite{Waldram1970}  More recent theoretical investigations have used the framework of the quasiclassical theory of superconductivity to discuss the characteristics of SNS junctions.\cite{Rammer1986}  Qualitatively, the physics of the SNS junction is well understood.   The proximity to the superconductor has two major effects on the quasiparticles in the normal metal.\cite{Pannetier2000}  First, it induces superconducting-like correlations between quasiparticles that increase the conductance of the normal metal, and second, it induces a gap in the quasiparticle density of states $N(E)$.\cite{leSueur2008}   While in the superconductor the energy scale for the density of states is given by $\Delta$, the energy scale for $N(E)$ in the normal metal for the diffusive case is set by the Thouless energy $E_{Th}= \hbar D/L^2$, where $D = v_F \ell /3$ is the quasiparticle diffusion constant in the normal metal, $\ell$ being the elastic scattering length, and $L$ is the length of the normal metal.   Similarly,  the maximal supercurrent that can flow through such a proximity junction is set by $E_{Th}$, not by $\Delta$ as for a conventional Josephson tunnel junction, when $E_{Th} \ll \Delta$ (the long junction limit).\cite{Wilhelm1997}  However, it is difficult to obtain quantitative predictions for the current-voltage characteristics except in limiting cases.  The problem arises from the fact that a finite voltage results in a time dependent phase difference between the superconducting electrodes, which in turn leads to time-dependent boundary conditions for the quasiclassical equations.  These equations have been solved under certain simplifying assumptions, such as low interface NS interface transparency,\cite{Bezuglyi1999} or in the limit of voltages small compared to $E_{Th}/e$.\cite{Brinkman2003}  Unfortunately, these assumptions do not apply to our samples.  Consequently, in the discussion below, we have chosen to discuss our results phenomenologically in the framework of  RSJ model discussed above, but with the energy potential landscape being calculated using the quasiclassical theory of superconductivity.  

The Josephson junction model leads to an interesting extension, where two junctions can be connected in parallel to form a dc superconducting quantum interference device (SQUID).  In this case, there are two independent parameters (the phase differences across the two junctions), giving rise to a two-dimensional potential for the system.  The system can transition from one local minimum to another through saddle-points in the potential. \cite{Tesche1977} Due to the fact that the position of the system on the two dimensional potential is sensitive to the external magnetic flux $\Phi$,  dc SQUIDs have been investigated extensively due to their device potential.\cite{Clarke2004,Hopkins2005}  With any real dc SQUID, the two Josephson junctions cannot be fabricated to be exactly the same, which restricts the ability to tune the system.

\begin{figure}
\includegraphics[width=8.5cm]{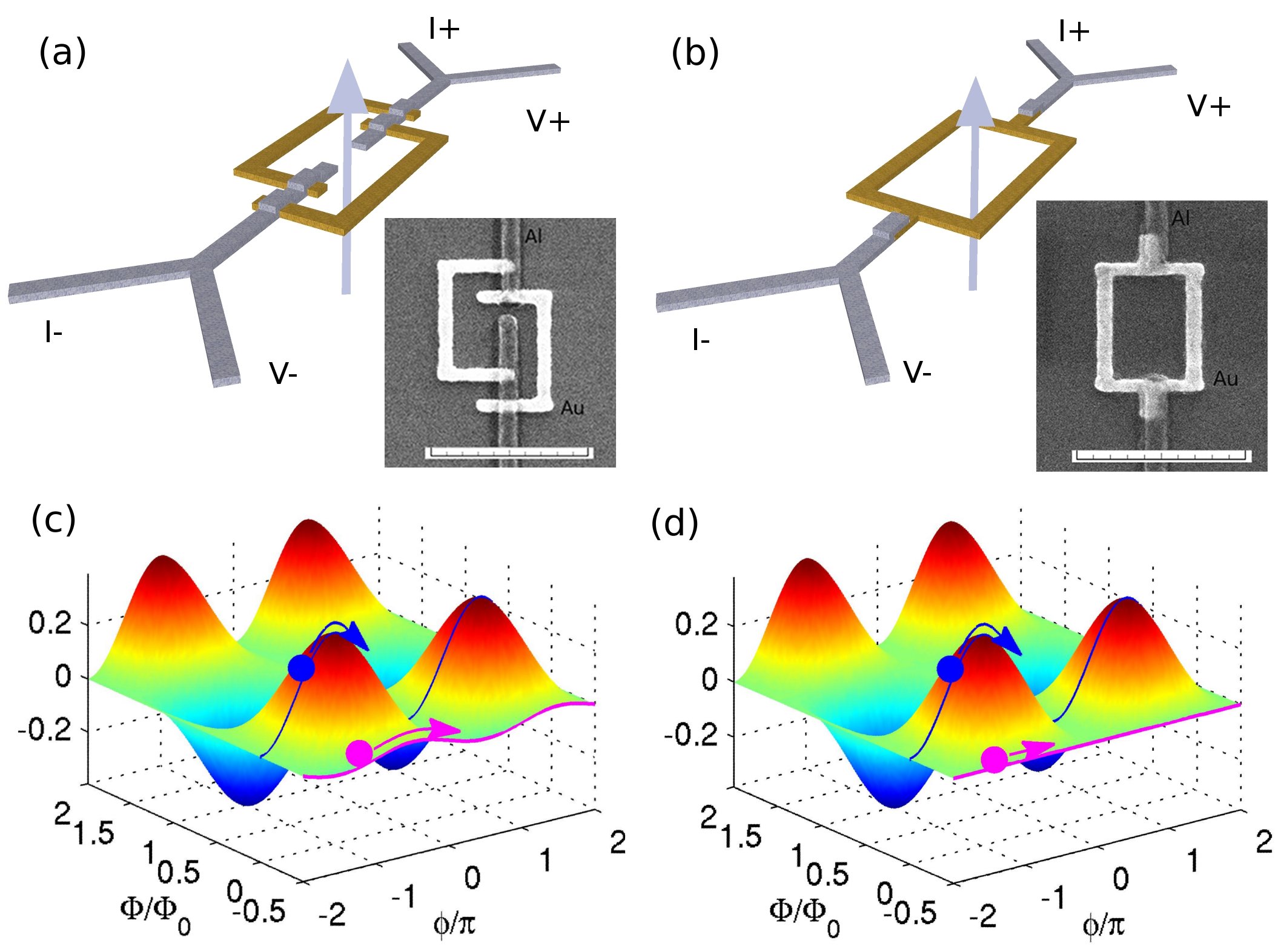}
\caption{(Color online) Schematic diagrams of two SNS quantum interference devices: asymmetric (a) and symmetric (b). The normal metal arms are shown in gold and the superconducting wires are shown in gray. The arrow in both figures corresponds to the the direction of applied magnetic flux. Insets: Scanning electron micrographs of the devices measured, the scale bars are 1 $\mu$m.  The calculated energy profile of the asymmetric/symmetric device are shown in (c)/(d) as a function of the external magnetic flux $\Phi$ and the phase difference across the two superconductors $\phi$, which is determined by the external current through the device. The energy profiles are calculated based on the quasiclassical theory, as described in the text. }  
\label{fig1}
\end{figure}

As with conventional junctions, two proximity junctions may be combined in parallel to form a dc SQUID.  Schematics of two types of SNS junctions are shown in Fig.~\ref{fig1}. In the asymmetric device shown in Fig.~1(a), the normal metal arms (shown in gold) connect to the superconducting wires (shown in gray) at different points, likely resulting in different NS interface transparencies for the two arms of the device, and consequently an asymmetric dc SQUID.\cite{Angers2008,Wei2008a} However, the long-range nature of the Josephson coupling in SNS devices enables a new type of device that is not possible with conventional tunnel junctions.  As shown schematically in Fig. \ref{fig1}(b), the device consists of a single normal metal loop between the two superconducting contacts, so that the NS interface transparencies for the two arms of the device are the same.  In this device, the modulation of quantum interference by an external magnetic flux occurs within the junction itself, i.e., the superconducting phase winding happens along the loop inside the junction, similar to the case of a superconducting loop that shows classical Little-Parks oscillations.\cite{Little1962,Moshchalkov1993,Liu2001}   As the NS interfaces are the same for both arms of the loop, this device potentially can behave as a perfectly symmetric dc SQUID, provided that the length of the two normal metal arms are the same.
 
To understand the flux-tunable thermal activation barrier in these SNS junctions, we use quasiclassical theory and a simple one dimensional model to calculate the energy of the system as a function of the phase difference $\varphi$ between the two superconductors (see appendix for details).  The energy of the system is given by (see, e.g., Ref.~[\onlinecite{Tinkham1996}], page 198) 
\begin{equation}
  E_J (\varphi, \Phi) = \frac{\hbar}{2e} \int^\varphi d\varphi ' I_s(\varphi ', \Phi),
\label{eqn_E_Phi}
\end{equation}          
where $I_s(\varphi ', \Phi)$ is the supercurrent through the system, which is a periodic function of the phase difference $\varphi '$ and the externally applied flux $\Phi$.  To calculate $I_s$, we use the extended circuit theory~\cite{Stoof1996} and numerically solve the Usadel equations for the sample geometries shown in Figs.~\ref{fig1}(a) and \ref{fig1}(b). 
In the simple model we assume that the interfaces between the normal metal (N) and the superconductors (S) are perfectly transparent, the gap $\Delta$ regains its bulk value in the superconductor within a very short distance of the NS interface, and that the distribution of quasiparticles in the superconducting reservoirs is given by the equilibrium Fermi function $f(E)$. Since the characteristic unit for the supercurrent for an SNS junction is $E_{Th} /eR$,  we use $(\hbar/2e^{2}R)E_{Th}$ as the characteristic unit of energy for $E_J$,\cite{Wilhelm1997} with $L$ being the length of one side of the loop.  For the parameters used in this simulation, the amplitude of the supercurrent is about 0.2 $E_{Th} /eR$, so the  modulation of $E_J$ at fixed $\Phi$ is about 0.4 $(\hbar/2e^{2}R)E_{Th}$  (note that since $E_J$ has an arbitrary constant from the integration in Eq.~(1), in Fig.~\ref{fig1} we assume $E_J =0$ at $\varphi =-2 \pi$). 

The resulting energy profiles are shown in Figs.~\ref{fig1}(c) and \ref{fig1}(d) as a function of $\varphi$ and $\Phi$, the two parameters under external experimental control.   If $\Phi$ is fixed at integral values of the superconducting flux quantum $\Phi_0=h/2e$, there is an energy barrier for evolution of the phase, as shown by the trajectories of the blue particles at zero flux, so that at low temperatures, the phase $\varphi$ is stationary, and no voltage is developed across the device.
For half-integral values of the applied flux ($\Phi=(n+1/2)\Phi_0$, where $n$ is an integer) there is a difference between the asymmetric and symmetric cases.  For the asymmetric case, there is still a small energy barrier, as shown by the trajectory of the red  particle in Fig.~\ref{fig1}(c).
 Consequently,  the resistance at odd half-integral flux quanta will eventually vanish if the temperature is low enough, in the absence of quantum tunnelling.  In contrast, for the symmetric case, there is no energy barrier at odd half-integral flux quanta (as shown by the trajectory of the red particle in Fig~\ref{fig1}(d)), so that the device will have a finite resistance even at the lowest temperatures.

\section{Zero bias magnetoresistance}

\begin{figure}
\includegraphics[width=8.5cm]{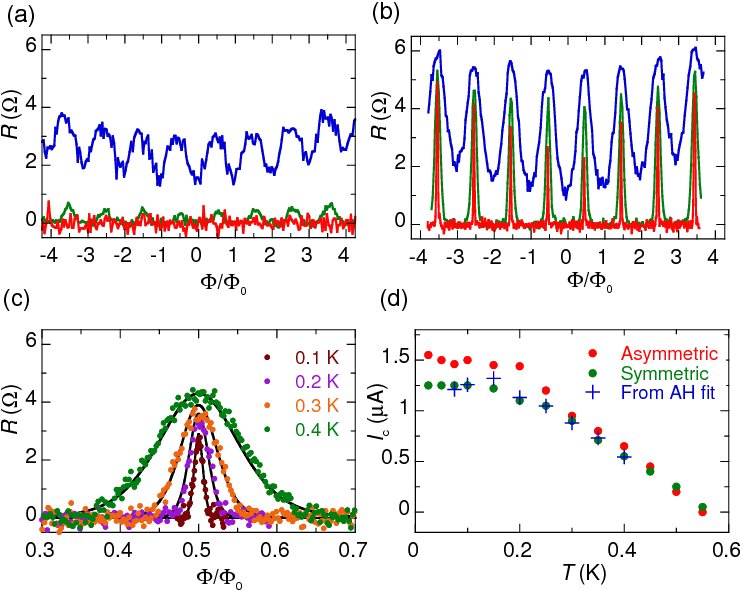}
\caption  {(Color online)  Resistance as a function of applied magnetic flux for the asymmetric sample (a) and the symmetric sample (b)  at 0.6  K (blue curves), 0.4 K (green curves), and 0.03 K (red curves).  While the oscillations for the asymmetric device die out rapidly with decreasing temperature, the oscillations for the symmetric device survive to the lowest measurement temperature.  (c)  Magnetoresistance of the symmetric device around $\Phi=\Phi_0/2$ at four different temperatures.  The solid lines are fits to the AH theory as described in the text.  Below 0.15 K, the magnetoresistance does not change with temperature.  (d)   Critical current at zero applied flux for the symmetric and asymmetric devices.  The plus symbols show the critical current expected from the fits of (c), multiplied by a factor of 5, as described in the text.} 
\label{fig2}
\end{figure}

Figures \ref{fig2}(a) and \ref{fig2}(b) show the resistance of the asymmetric and symmetric devices respectively as a function of applied magnetic flux at a number of temperatures. The details of fabrication and measurement are similar to those reported elsewhere.~\cite{Wei2008a}
The data are taken in the limit of zero dc current, with only a very small ac current for the resistance measurement (about 10-20 nA).  

At higher temperatures, the resistance is finite at all values of $\Phi$ for both geometries,  and is periodic in the applied flux, with a fundamental period of $\Phi_0$.  As the temperature is lowered, the resistance for both devices vanishes near $\Phi=n\Phi_0$.  As the temperature is lowered still further, the resistance of the asymmetric device also vanishes at odd half-integral values $\Phi=(n+1/2)\Phi_0$ of the applied flux.  In contrast, the resistance of the symmetric device at $\Phi=(n+1/2)\Phi_0$ remains finite down to the lowest temperatures, while the peak width narrows as the temperature is lowered, saturating below about 0.15 K, as shown for the peak around $\Phi=\Phi_0/2$ in Fig.~\ref{fig2}(c).

According to the physical picture for the symmetric device in Fig.~\ref{fig1}(d), there is a small but finite barrier for the system to overcome, except at exactly half-integral values of flux.  This barrier decreases monotonically as the system approaches $\Phi=(n+1/2)\Phi_0$. In fact, for conventional junctions the Josephson coupling energy is  $E_j = (\hbar/2e) I_c$, $I_c$ being the critical current.  In our SNS junction case, $E_j$ is given by Eq.~(\ref{eqn_E_Phi}), and is a function of the external flux $\Phi$.   Assuming symmetrical long junctions, we obtain the usual sinusoidal dependence of the current $I_s$ on $\varphi$, with the Josephson coupling energy
\begin{equation}
E_j = \frac{\hbar I_c(\Phi)}{2e},
\label{eqn_E_j}
\end{equation}
where $I_c(\Phi) = I_c(0) |\cos (\pi\Phi/\Phi_0)|$, the same as that for a symmetric dc SQUID (see, e.g., Ref.~[\onlinecite{Tinkham1996}], page 215).
 Thus, as the temperature is lowered, the system needs to be closer to $\Phi=(n+1/2)\Phi_0$ until the energy barrier is low enough for the particle to jump over and for the junction to exhibit a finite resistance.  At even lower temperatures, the system may tunnel through the barrier, resulting in a temperature independent resistance. However, it is necessary to verify whether the electron temperature follows the bath temperature or not,\cite{Courtois2008} which is non-trivial and will not be discussed here. We note that the measured critical current saturates at lower temperatures and is about 10 times smaller than the value predicted in the zero temperature limit,\cite{Wilhelm1997,Angers2008} probably due to an imperfect SN interface, fluctuations, and heating.\cite{Courtois2008}

To quantitatively model the magnetoresistance around half-integral flux values for the symmetric device, we can use AH theory for RSJ model, as extended here to a dc-SQUID with a flux-tunable barrier.  The AH theory predicts that the normalized resistance in the zero current limit is given by 
\begin{equation}
  R_{AH}= \mathcal{I}_0^{-2} (\gamma/2)
\label{eqn_gamma}
\end{equation}
where $\mathcal{I}_0$ is the modified Bessel function, and $\gamma=2E_j/k_B T$ is the ratio between the barrier and the thermal fluctuation energy.   From Eq.~(\ref{eqn_E_j}), we have $\gamma = \hbar I_c(\Phi) /e k_B T$, and we can then fit the magnetoresistance of the symmetric device near $\Phi=\Phi_0/2$ using the AH theory with Eq.~(\ref{eqn_gamma}).  

The solid lines in Fig.~\ref{fig2}(c) show the resulting fits of $R(\Phi)$ at four different bath temperatures, using only the measured peak resistance $R_p$ at $\Phi=\Phi_0/2$ and zero field critical current $I_c(0)$ as fitting parameters. Figure~\ref{fig2}(d) shows a comparison of $I_c(0)$ obtained from the fits (multiplied by a factor of 5) compared to the experimentally measured values of $I_c(0)$. The fitted values of $I_c(0)$ are a factor of 5 smaller than the experimentally measured values of $I_c(0)$ over the entire temperature range. Conversely, we can claim that the effective noise temperature $T_N$ is 5 times larger than $T_b$ since only the ratio between $E_j$ and $k_B T$ matters.  

The strong  correlation between the measured and the fitted values indicates that the AH theory is somewhat applicable for the flux-tunable proximity junction near $\Phi=\Phi_0/2$, but the effective potential barrier for thermal activation appears to be smaller by about a factor of 5 compared to Eq.~(\ref{eqn_E_j}). As this factor is temperature independent, this ``enhanced'' fluctuation could not be due to quantum fluctuations,\cite{Liu2001} nor due to heating. In fact, if there is heating during measurements of $I_c$ then the measured $I_c$ should be smaller than that inferred from the thermal activation model at zero current limit, but the opposite is observed.  This discrepancy could be related to the assumption of a standard external shunt resistor as the thermal noise source in the RSJ model, while the SNS junction is self-shunted, or it could be due to the neglect of inverse proximity effect in superconducting electrodes and formation of minigap in the normal metal,~\cite{leSueur2008} which may change the relation between $I_c$ and $E_j$ in Eq.~(\ref{eqn_E_j}). To better understand this ``enhanced'' fluctuation, below we characterize devices beyond the zero bias limit.

\section{Finite bias}

\begin{figure}
\includegraphics[width=8.5cm]{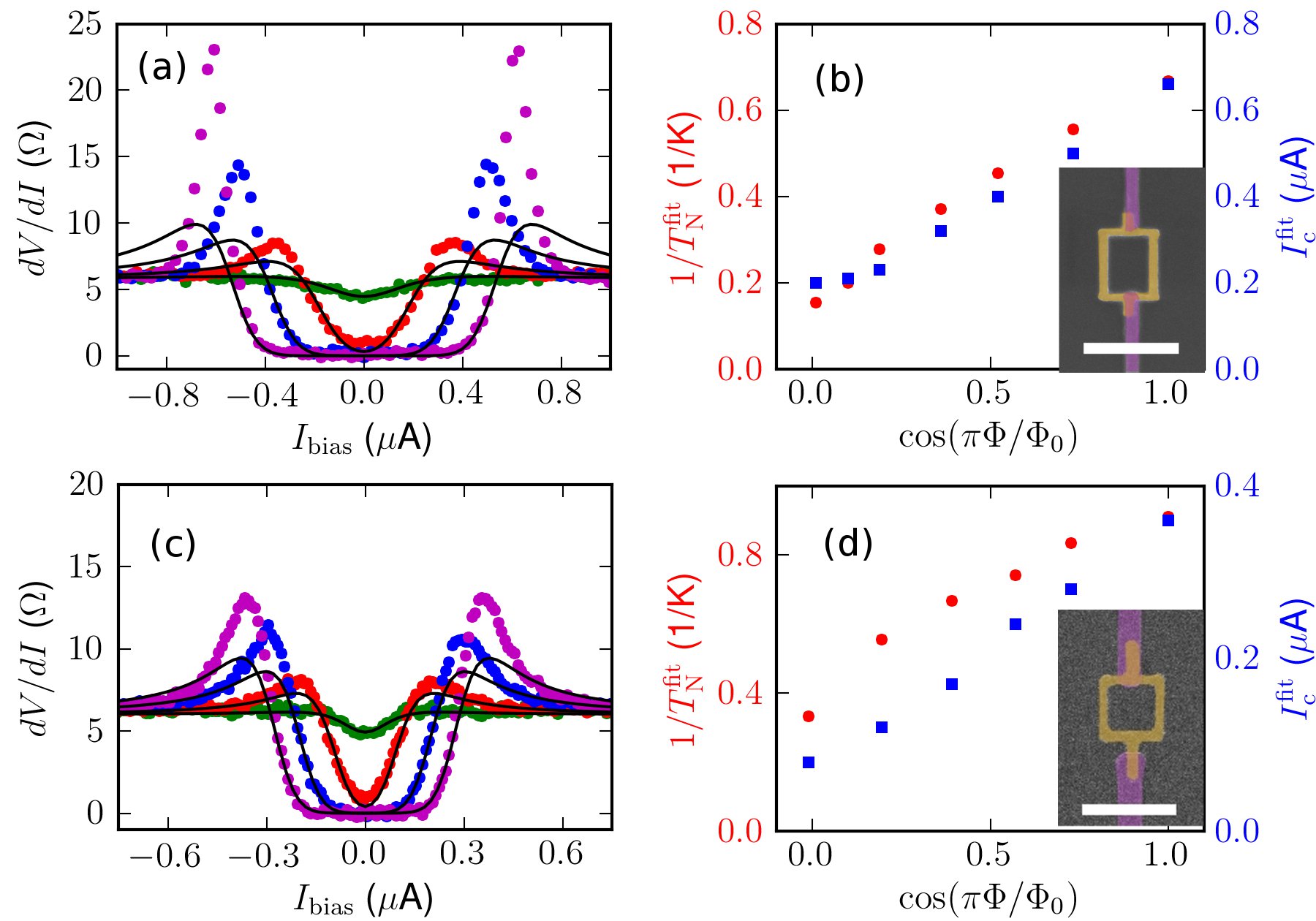}
\caption{(Color online) Differential resistance at 0.4 K for $s1$, (a), and for $s2$, (c). Different colors correspond to different applied magnetic flux: 0 (magenta), 0.25 (blue), 0.375 (red), and 0.5 (green) in units of $\Phi_{0}$.  Solid lines are fits using AH model. (b) and (d) show the flux dependence of the noise temperature $T_{N}$ (red) and critical current $I_{c}$ (blue) extracted from the AH fits of the differential resistance in (a) and (c). The insets are SEM images of the devices with false color enhancement. The area colored in gold is Au and the area colored in purple is Al. The white scale bars are 1 $\mu$m.}  
\label{fig_dVdI_AH}
\end{figure}

Finite bias measurements were conducted on two representative proximity junction devices with superconducting Al electrodes of different widths: about 110 nm and 220 nm respectively for devices $s1$ and $s2$, see Table~\ref{table_sample} for other parameters, and insets of Fig.~\ref{fig_dVdI_AH} for SEM images. The Al wires extend for several micrometers on each side of the loop before overlapping with Au leads for four-probe measurements.

\begin{table}
\caption{Sample parameters of two representative proximity junction devices: w$_{Al}$ and t$_{Al}$ are the width and thickness of the Al electrodes, w$_{Au}$ and t$_{Au}$ are the width and thickness of the Au loops, T$_{c}$ and H$_{c}$ are the critical temperature and critical magnetic field of the Al electrodes, which is sensitive to w$_{Al}$ for quasi-1D wires. }
 \begin{tabular*}{8.5cm}{@{\extracolsep{\fill}}ccccccc}
\hline\hline
\ & w$_{Al}$ & t$_{Al}$ & w$_{Au}$ & t$_{Au}$ & T$_{c}$ & H$_{c}$ \\
  &  nm  & nm &  nm  & nm & K & Gauss \\
  \hline
$s1$ & 110 & 88 & 90 & 50 & 1.2 & 600 \\
$s2$ & 220 & 81 & 90 & 45 & 1.1 & 220 \\
\hline\hline
 \end{tabular*} 
\label{table_sample}
\end{table} 

As in early investigations for conventional Josephson junctions,~\cite{Anderson1969,Simmonds1970,Falco1974} we measured the differential resistance of the two proximity junction devices at 0.4 K, as shown in Fig.~\ref{fig_dVdI_AH}, and fit the data by numerical differentiation of the voltage in the AH theory.\cite{Ambegaokar1969}  In the limit of small currents ($x=I/I_{C}<1$) and low temperatures ($\gamma=I_{C}\hbar/ek_{B}T \gg 1$), the normalized voltage $v=V/I_{C}R$ is
\begin{equation}
 v=2\sqrt{1-x^{2}}\exp[-\gamma[\sqrt{1-x^{2}}+x\sin^{-1}x]]\sinh[\pi\gamma x/2].
\label{eqn_v_AH}
\end{equation} 
As for $R_{AH}$ in Eq.~(\ref{eqn_gamma}), the essential fitting parameter is $\gamma$, the ratio between the energy barrier $E_{J}$ and the thermal energy $k_{B}T_b$. Assuming $E_j$ is just the Josephson coupling energy in Eq.~(\ref{eqn_E_j}), then we need to replace $T_b$ (0.4 K) with an effective noise temperature $T_{N}$ to characterize the enhancement of fluctuations.  

In Fig.~\ref{fig_dVdI_AH} we show the result of fitting at several different flux values (in unit of $\Phi_{o}$), with both $T_{N}$ and $I_c$ used as fitting parameters.  When $E_J$ is suppressed at around half flux quanta, $T_{N}$ reaches its maximum, and the quality of the fit is good to large values of $I$.  When $E_J$ is maximum at around integer flux quanta, there is clear deviation of the fit from the experimental data as the bias current approaches $I_c$, and we can only fit the low current bias regime. Compared to previous investigations,\cite{Anderson1969,Simmonds1970,Falco1974} here  we can tune the Josephson coupling ($E_J$) without varying the physical temperature, which gives us a knob to tune $\gamma$ at a constant temperature.  

\begin{table*}
\caption{Fitting parameters. See Fig.~\ref{fig_dVdI_AH} for fitting of dV/dI using AH theory, Fig.~\ref{fig_IVC_AH} for fitting of IVC using AH theory, Fig.~\ref{fig_IVC_LAMH} for fitting of IVC using LAMH theory, and Fig.~\ref{fig_RT_fit} for fitting of $R(T)$ with three different models.  }
 \begin{tabular*}{1.\textwidth}{@{\extracolsep{\fill}}c|c|c|ccc|ccc|cccc|*{4}{c}} 
\hline\hline
\ & dV/dI AH fit & IVC AH fit & \multicolumn{3}{c|}{IVC LAMH fit} & 
\multicolumn{3}{c|}{R(T) Little fit} & \multicolumn{4}{c|}{R(T) LAMH fit} & \multicolumn{4}{c}{R(T) SNS fit} \\

 & $(dV/dI)_{N}$ & R$_{N}$ & I$_{0}$ & A & B & R$_{N}$ & T$_{c}$ & c &
 R$_{N}$ & T$_{c}$ & D & c & R$_{N}$ &  T$_{c}$ & a & b \\
 & $\Omega$ &$\Omega$ & $\mu$A & $\mu$V &  & $\Omega$ & K &  &
$\Omega$ & K & $\Omega$ &  & $\Omega$ &  K &  &  \\

\hline
$s1$ & 5.8 & 5.8 & 0.07 & 0.33 & 6.2 & 6 & 0.78 & 14 & 6 & 0.92 & 550 & 15 & 6 & 1.2 & $1\times10^{5}$ & 14.8 \\
\hline
$s2$ & 6 & 5.8 & 0.04 & 0.11 & 5.8 & 6 & 0.64 & 14 & 6 & 0.77 & 550 & 15 & 5.5 & 1.1 & $7\times10^{4}$ & 15 \\ 
\hline\hline
  
 \end{tabular*}
\label{table_fit}
\end{table*} 

In Fig.~\ref{fig_dVdI_AH} (b) and (d), the values of $I_c$ and $T_{N}$ resulting from the fits are plotted as a function of $\cos(\pi \Phi/\Phi_{0})$, i.e., the normalized $E_j$.  $I_c$ has a linear dependence on $\cos(\pi \Phi/\Phi_{0})$, as expected. $T_{N}$ also increases monotonically from $\Phi=\Phi_0$ to $\Phi = 0$, but the dependence of $T_{N}$ with flux is difficult to interpret within the AH theory.  For conventional SQUIDs, it is known that near half flux quanta there are multiple metastable states which may lead to enhanced fluctuations. However, such enhanced fluctuations have not been reported in previous escape rate experiments.\cite{Goldman1965,Naor1982,Sharifi1988,Han1989,Lefevre-Seguin1992} A somewhat similar  scenario is the mesoscopic fluctuations in SNS junctions considering the situation that several stationary states  may exist at a given current,\cite{Altshuler1987,Hoss2000,Houzet2008} but the supercurrent fluctuation cannot explain the flux dependence.

It is clear that the fitted values of $T_{N}$ are larger for $s1$ than for $s2$. Since $s1$ has thinner superconducting leads, a larger $T_{N}$ may suggest that the thermal activation energy barrier is related to the dimensions of the superconducting leads, reminiscent of the fact that in TAPS theory $\Delta F$ is proportional to the cross-sectional area. Further investigation is required to verify this claim.

\begin{figure}
\includegraphics[width=8.5cm]{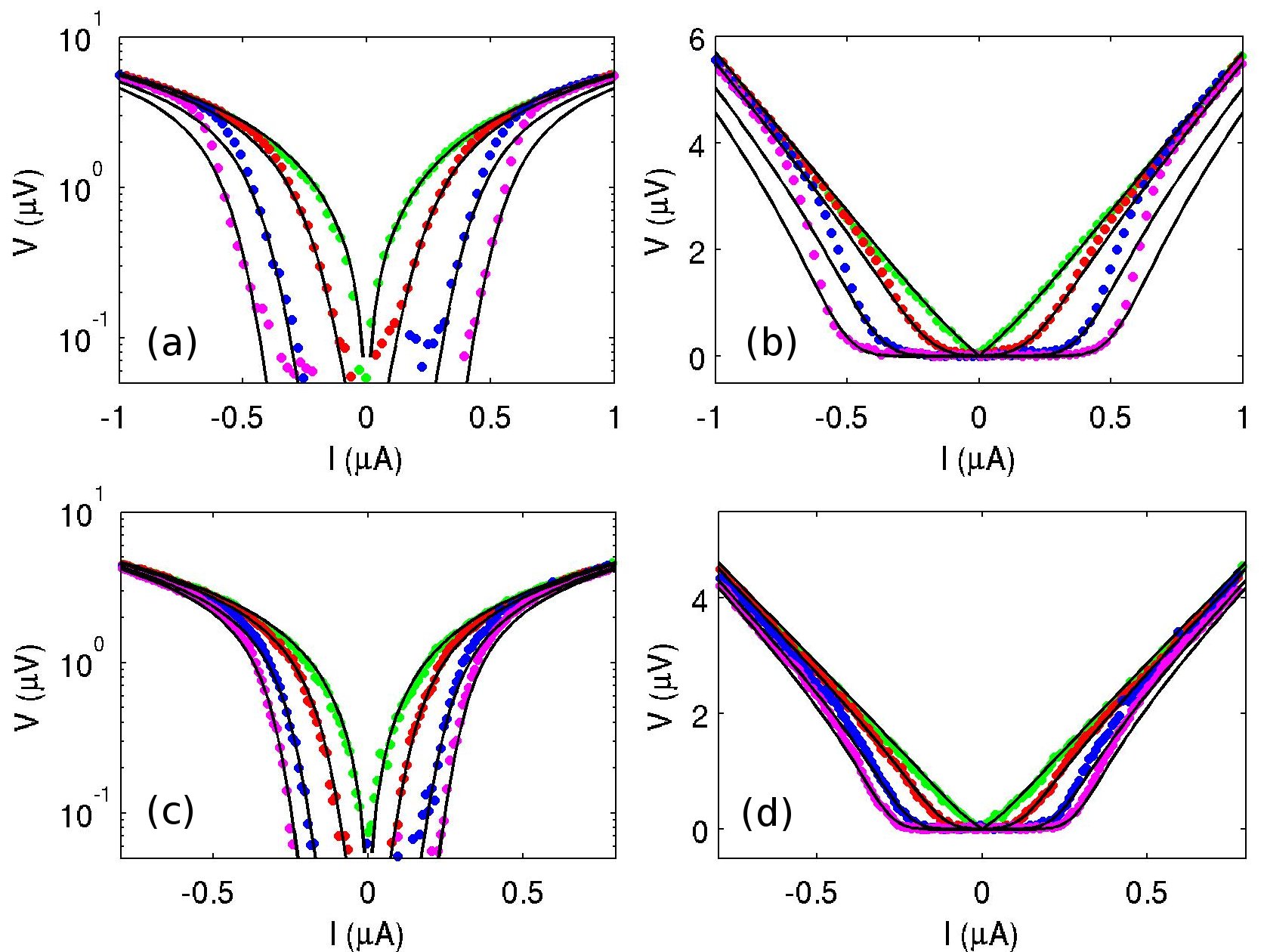}
\caption{(Color online) Current voltage characteristics at 0.4 K for $s1$: (a) and (b), and $s2$: (c) and (d). The absolute value of voltage is plotted as the y axis in (a) and (c), in log scale, and in (b) and (d), in linear scale. Solid lines are fits using AH theory with the same $T_{N}$ in Fig.~\ref{fig_dVdI_AH}. }  
\label{fig_IVC_AH}
\end{figure}

In recent experimental investigations  on proximity junctions,\cite{Hoss2000,Dubos2001b,Lhotel2007} current-voltage characteristics (CVC) were often presented, and strong broadening was reported when comparing experimental data to the intrinsic broadening predicted by the AH theory and RSJ model. Here the measured CVC (measured at the same time as the differential resistance) are plotted in Fig.~\ref{fig_IVC_AH} for comparison.  The strong broadening can be fit well in the regime $I<I_{c}$ by Eq.~(\ref{eqn_v_AH})  with the same $T_{N}$ in Fig.~\ref{fig_dVdI_AH}. At higher current bias the CVC collapse to a single curve, and cannot be fit. For sample $s2$, although the superconducting electrodes are wider than that of $s1$, the critical current is smaller probably due to a less transparent interface. The deviation at higher bias is less obvious for $s2$ since the spreading of $I_c$ is smaller.  We also show in Fig. 4 the CVC on a logarithmic scale for later comparison with fits using the TAPS model.

\begin{figure}
\includegraphics[width=8.5cm]{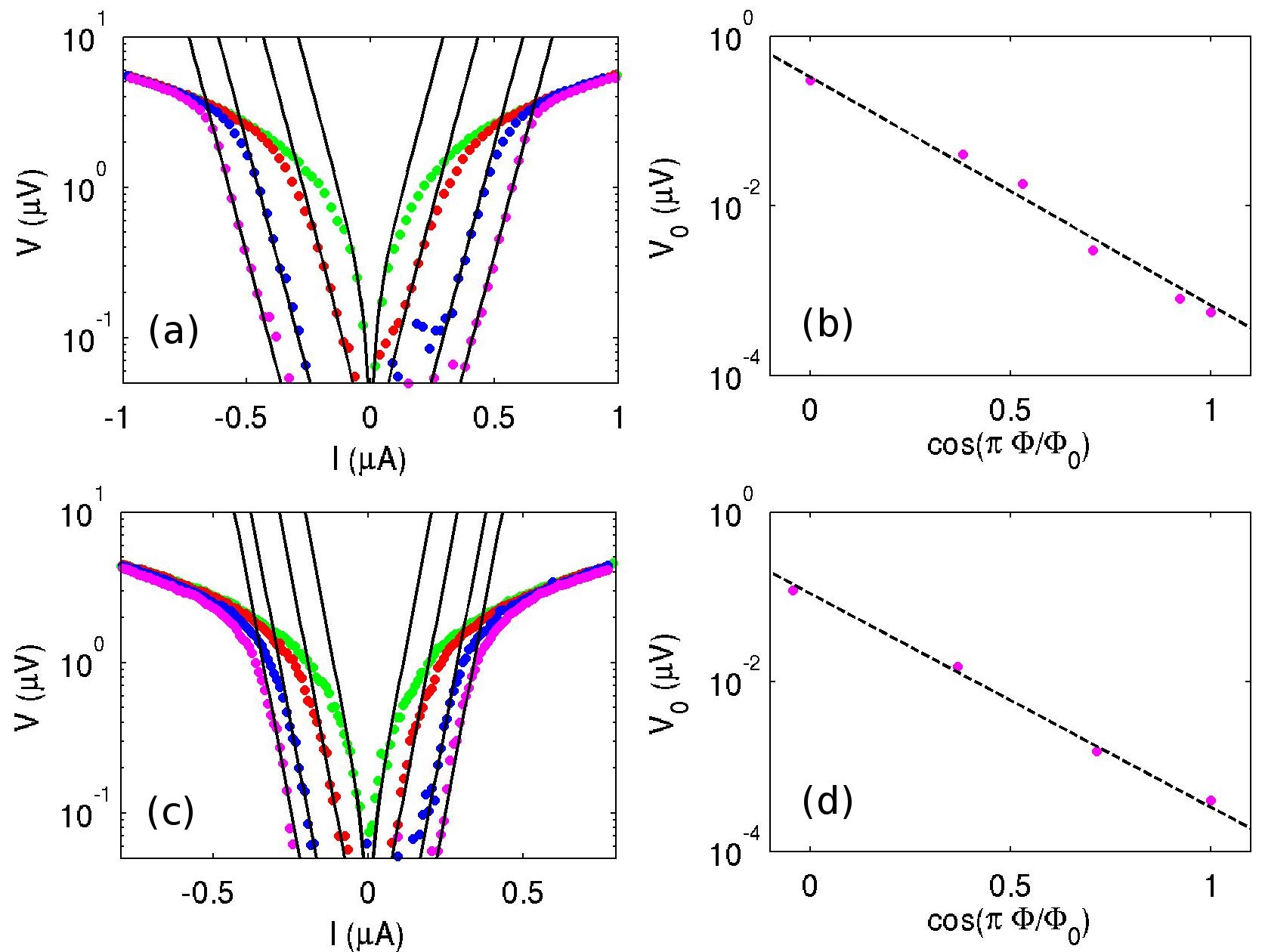}
\caption{(Color online) Current voltage characteristics at 0.4 K for $s1$, (a), and for $s2$, (c), fitted by $V(I)=V_{0}\sinh(I/I_{0})$. In (b) and (d) the fitting parameter $V_{0}$ is plotted as a function of flux. }  
\label{fig_IVC_LAMH}
\end{figure}

As discussed in the introduction, in terms of analysis, superconducting proximity junctions lie somewhere between weakly coupled conventional junctions and (strongly coupled) superconducting wires.  Consequently, besides the AH theory for junctions, we also can try the LAMH theory for superconducting wires to fit the CVC of our devices. For thin superconducting wires, the average voltage generated by TAPS due to thermal fluctuations is~\cite{Langer1967,McCumber1970}
\begin{equation}
V_{LAMH}=\frac{\hbar\Omega}{e}e^{-\Delta F_{0}/kT}\sinh\frac{\delta F}{2kT},
\label{eqn_V_LAMH}
\end{equation}
where $\Delta F_{0} = (8\sqrt{2}/3)(H_{c}^{2}/8\pi)A\xi$ is the energy barrier, $\Omega = (L/\xi)(\Delta F_{0}/kT)^{1/2}\tau^{-1}_{GL}$ is the attempt frequency, $\tau_{GL}=\pi\hbar/8k(T_{c}-T)$ is the Ginzburg-Landau (GL)  relaxation time, $\delta F = hI/2e$ is the difference in the energy barrier for phase slips in two directions, $A$ is the cross-sectional area, and $\xi$ is the GL coherence length. Note that the hyperbolic sine term in Eq.~(\ref{eqn_V_LAMH}) has the same form as in the AH theory (see Eq.~(\ref{eqn_v_AH})). For our devices, $\Delta F_{0}$ should be considerably lower than $\Delta F_{0}$ for a superconducting wire .  
Using interrelations of quantities from GL theory, the energy barrier can be reformulated as~\cite{Tinkham2002} 
\begin{equation}
\Delta F_{0} = \sqrt{6}(\hbar/2e)I_{c},
\label{eqn_Delta_F}
\end{equation}
which is similar to the form of $E_J$ in Eq. (2).  Thus,  $\Delta F_{0}/kT$ in the LAMH theory is comparable to $\gamma$ in the AH theory.

At constant temperature, Eq.~(\ref{eqn_V_LAMH}) can be simplified to $V_{LAMH}(I)=V_{0}\sinh(I/I_{0})$,  as $\Omega$ and $\Delta F_{0}$  should have a much weaker dependence on the bias current $I$ than the hyperbolic sine term.\cite{Rogachev2005} Here $V_{0}$ is the flux dependent prefactor, and $I_{0}=4ek_{B}T/h$ is a constant~\cite{Tinkham1996} at fixed temperature. At 0.4 K  $I_{0}$ is $0.0052$ $\mu A$. In Fig.~\ref{fig_IVC_LAMH} CVC at different fields are fitted by this simplified equation (see Table~\ref{table_fit} for fitting parameters). The fitting parameter $I_{0}$  for $s1$ ($s2$) is $\sim 0.07$ (0.04) $\mu A$, about 10 times larger than the expected value.  Previously, about 40\% increase of $I_{0}$ was reported for MoGe superconducting nanowires,\cite{Rogachev2005} but here the increase of $I_{0}$ is much larger. To characterize the increase of $I_{0}$, we can again assume an effective noise temperature $T_{N}$ that is about 10 times higher than the bath temperature $T_b$, similar to that in fitting with AH theory. However, here $T_{N}$ is flux independent, in contrast to the AH fit. 

The constant $V_{0}$ can be fit with a simple exponential equation $y=A e^{-Bx}$ as shown in (b) and (d) of Fig.~\ref{fig_IVC_LAMH}, following Eq.~(\ref{eqn_V_LAMH}). Using Eq.~(\ref{eqn_Delta_F}) and $I_c(\Phi) = I_c(0) |\cos (\pi\Phi/\Phi_0)|$, we can rewrite $V_{0}$ as 
\begin{equation}
V_{0}(x)=Ae^{-Bx}=\frac{\hbar\Omega}{e}\exp[-\frac{\sqrt{6}\hbar I_{c}(\Phi=0)}{2ekT}x],
\label{eqn_V_LAMH2}
\end{equation}
where $x=\cos(\Phi/\Phi_{0})$. The factor in front of $x$ can be reformulated as $B=\sqrt{6}I_{c}(\Phi=0)/\pi I_{0}$. In  Fig.~\ref{fig_IVC_LAMH}~(a), if we define the critical current $I_c$ as the current at which the measured voltage approaches 1 $\mu$V, then  at zero flux $I_{c} \sim 0.56$ $\mu A$. Since $I_{0}=0.07$ $\mu$A for $s1$,  the expected $B \sim 6.24$, very close to the fitted value of $B_{fit}=6.2$ (see Table~\ref{table_fit}). This means we do not need to introduce parameters other than the flux independent $T_{N}$ to fit all CVC data. For $s2$ the expected $B$ is 6.83, slightly higher than $B_{fit}=5.8$, which could be due some uncertainty in defining the $I_c$ values. Our analysis indicates that in general, it seems that the LAMH theory is better for fitting the CVC for proximity junctions.

\section{Resistance tail}

\begin{figure}
\includegraphics[width=8.5cm]{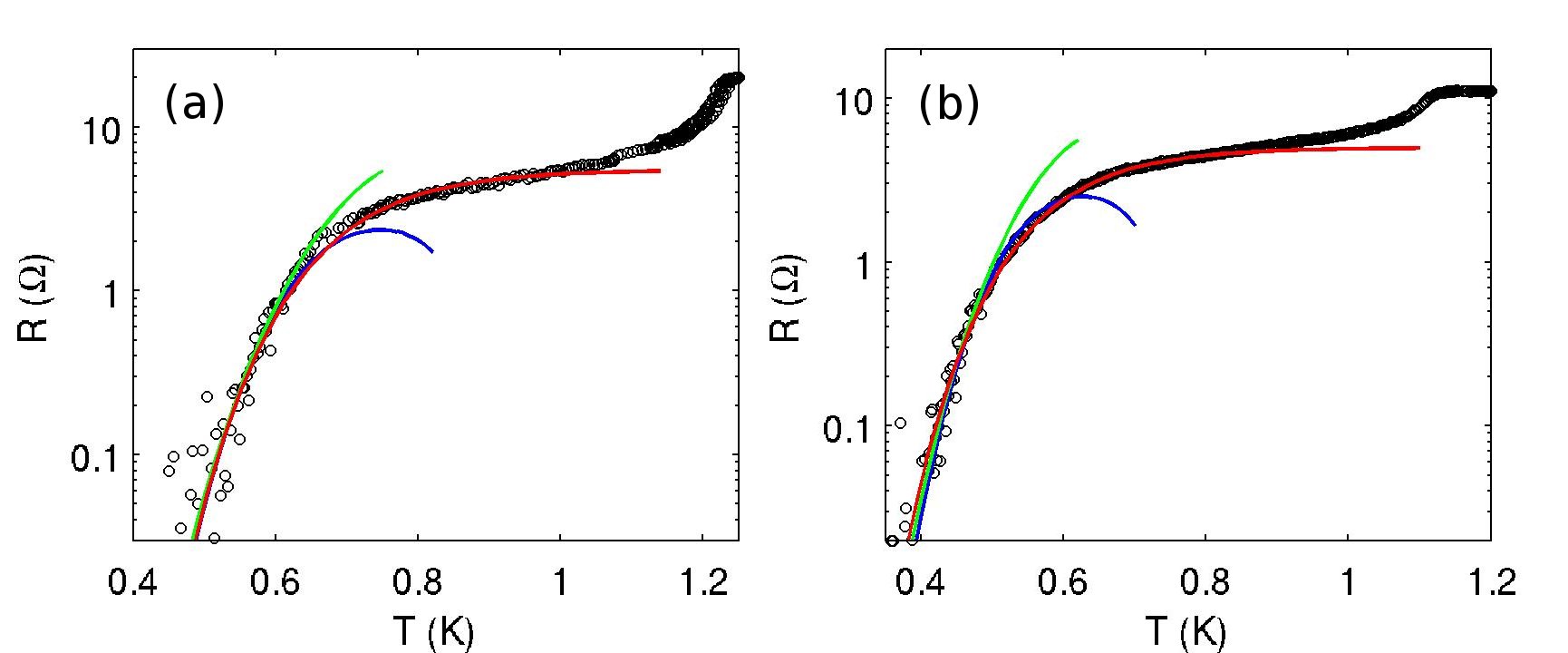}
\caption{(Color online) Resistance tail at zero field for $s1$, (a), and for $s2$, (b). The green lines are fits using Little's theory, the blue lines are fits using LAMH theory, and the red lines are fits combining Little's equation and quasiclassical theory for proximity junctions ($R_{SNS}(T)$). The fitting parameters are listed in Table~\ref{table_fit}.}  
\label{fig_RT_fit}
\end{figure}


The thermal activation theory  can fit the resistance tail slightly below $T_c$ for a thin superconducting wire. In the simplest case, we have the Arrhenius law proposed by Little\cite{Little1967,Bezryadin2008}
\begin{eqnarray}
R(T) &= & R_{N} e^{-\Delta F/k_{B}T}\nonumber \\
	& = & R_{N} \exp[-c(1-t)^{3/2}/t],
\label{eqn_RT_Little}
\end{eqnarray}
where $t=T/T_{c}$, $c = \Delta F_{0}(0)/kT_{c}$ is the temperature independent activation constant. 

As we noted earlier, the energy barrier for a proximity junction could be considerably lower than that for a superconducting wire, and Eq.~(\ref{eqn_Delta_F}) is used for a practical estimate of $c$ from the measured $I_c$. Taking  $B=\sqrt{6}I_{c}/\pi I_{0} \sim 6.24$ obtained at 0.4 K, and the fitting parameter $T_{c}=0.78 K$  (see Table~\ref{table_fit} for the fitted $T_c$ values with Little and LAMH models, which is somewhat arbitrary~\cite{Bezryadin2008}), from $B=c(1-t)^{3/2}/t$ we find  $c=9.41$, close to the $c_{fit}$ extracted from the R(T), as listed in Table~\ref{table_fit}. 

In the full LAMH theory the attempt frequency term is added to Eq.~(\ref{eqn_RT_Little}) (while the prefactor $R_{N}$ is dropped), and the resistance below $T_{c}$ derived from Eq.~(\ref{eqn_V_LAMH}) in the zero bias limit is~\cite{Bezryadin2008}
\begin{eqnarray}
R_{LAMH} & = & \frac{V_{0}}{I_{0}}e^{-\Delta F_{0}/kT} \nonumber \\
& = & Dt^{-3/2}(1-t)^{9/4}\exp[-c(1-t)^{3/2}/t],  \label{eqn_R_LAMH}
\end{eqnarray}
where $D$ is the attempt frequency constant. The total resistance is given by $R=(R^{-1}_{N}+R^{-1}_{LAMH})^{-1}$ as quasiparticles provide a parallel conduction channel to the supeconducting channel. The values of $c$ for the LAMH fits and and the fits to Eq. (8) are very close since $R(T)$ is very sensitive to the exponential term. The attempt frequency constant $D=(8/\pi)(L/\xi(0))R_{Q}\sqrt{c}$ is formulated for a superconducting wire with length $L$, where $R_{Q}\equiv h/4e^{2}$ is the resistance quantum. From the fitted values of $D$ and $c$ we get $L/\xi(0)\sim 0.01$, a very small value as expected for proximity junctions.   

In the analysis above, the fit for $R(T)$ is valid only in a narrow regime below the fitted $T_c$, which is due to the fact that $\Delta F_{0}$ is estimated using Eq.~(\ref{eqn_Delta_F}), where for superconducting wires near $T_c$, $I_c  \propto (1-t)^{3/2}$.  For SNS junctions, quasiclassical theory has shown that in the long junction limit ($\Delta \gg E_{Th}$), the simplified solution is\cite{Zaikin1981,Dubos2001a}
\begin{equation}
 eR_{N}I_{c}= (32/(3+2\sqrt{2}))E_{Th}(L/L_{T})^{3}e^{-L/L_{T}},
\label{eqn_Ic_QC}
\end{equation}
where  $L_{T}=\sqrt{\hbar D/2\pi k_{B}T}$ is the thermal length in the diffusive limit, and $D$ is the diffusion constant of the normal metal. We can put this $I_c(t)$ dependence into $\Delta F$ of Eq. (8), which results in:
\begin{eqnarray}
 R_{SNS}  =  R_{N}\exp[a\sqrt{t}e^{-b\sqrt{t}}], \nonumber \\
a=\frac{\hbar}{e^{2}R_{N}}\frac{16}{3+2\sqrt{2}}\pi b, \quad b=\frac{L}{\sqrt{\hbar D/2\pi k_{B} T_{c}}}.
\label{eqn_R_QC}
\end{eqnarray}

The formula is valid to much higher temperatures as shown by the red lines in Fig.~\ref{fig_RT_fit}, and there is no need to assume an arbitrary $T_c$ for the junction that is lower than $T_c$ of Al. Close to the $T_c$ of Al the fits deviate slightly from the data, probably because the long junction limit is no longer applicable, and also may due to the inverse proximity effect. With sample parameters $L\sim$ 1 $\mu$m, and $D\sim 100$ cm$^{2}$/sec, $a$ and $b$ are estimated to about $5\times10^{4}$ and $10$, the same order as the fitted values. We note that with the other thermal activation model, similarly fits can be achieved by using the zero current limit of junction resistance in AH theory  $R_{AH}  =  R_{N}\mathcal{I}_{0}^{-2}(\gamma/2)$, where $\mathcal{I}_{0}$ is the modified Bessel function. However, in that case the fitted $a$ values are much smaller than the estimate. 
       
\section{Summary}       
         
In conclusion,  we have measured nanoscale proximity junctions made of a mesoscopic normal-metal loop contacted by thin superconducting electrodes. The effect of thermal fluctuations in these devices can be characterized by an effective noise temperature. We find that the measurement results can be better described by the LAMH theory for thin superconducting wires, rather than by the AH theory for weak coupled Josephson junctions. With the LAMH theory, only the effective enhanced noise temperature (equivalent to the reduction of thermal activation barrier) is required to fit the data at different flux values.  We also find that quasiclassical theory can be combined with thermal activation theory to fit R(T) of proximity junction devices. These observations indicate that  for nanoscale proximity junctions we need to consider the finite size effect of the superconducting electrodes, and a quasiclassical model more completed than the simple one-dimensional model presented here may be needed to compute the reduction of the energy barrier. 

This research was conducted with support from the National Science Foundation under grant No. DMR-1006445.  

\appendix*
\section{Simulation}
 
The simulations shown in Fig.1 were done by solving the linearized Usadel equations in the Riccati parameterization.\cite{Belzig1999,Hammer2007}  In the normal-metal wire, the simplified Usadel equation reads
\begin{equation}
  \partial_{x}^{2}\gamma + 2i\epsilon \gamma= 0.
\end{equation}
Here the coherent function $\gamma$ is a complex function of distance $x$ and energy $\epsilon$, normalized by the total length $L$ and Thouless energy $E_{Th}$ respectively. The equation can be readily solved with appropriate boundary conditions. 

\begin{figure}
\includegraphics[width=8.5cm]{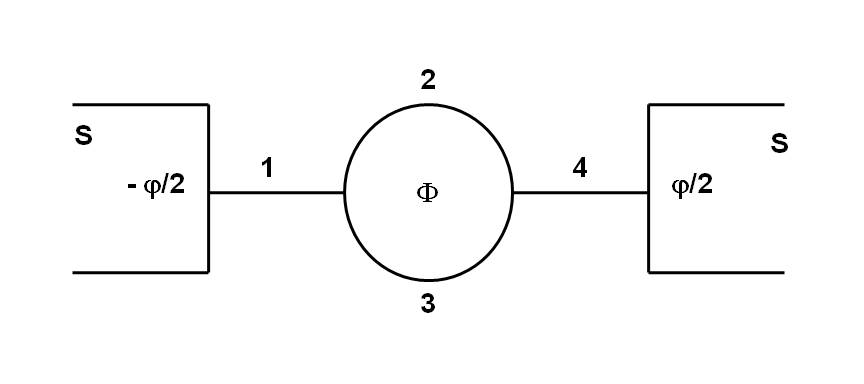}
\caption{Schematic of the SNS SQUID with applied magnetic flux $\Phi$ and phase difference $\varphi$ between two superconducting reservoirs. Numbers 1-4 correspond to normal-metal wire segments.} 
\label{figs1}
\end{figure}

For the SNS SQUID model shown in Fig.~\ref{figs1}, we match the solutions of 4 normal-metal wires by specifying boundary conditions at the two nodes and at the SN interface,\cite{Stoof1996,Golubov1997} e.g., at the node connecting wire 1, 2, and 3 we have
\begin{eqnarray}
\gamma_{r}(1) & = & \gamma_{l}(2),\nonumber \\
\gamma_{r}(1) & = & \gamma_{l}(3), \nonumber \\
A(1)\partial_{x}\gamma_{r}(1) & = & A(2)\partial_{x}\gamma_{l}(2)+A(3)\partial_{x}\gamma_{l}(3),
\end{eqnarray}
where $A(n)$ is the normalized product of the cross-sectional area and the normal conductance of the wire, and the subscript $l,r$ indicates the left or right end of the wire. Similar boundary conditions are used for the node connecting wire 2, 3, and 4 with phase factor due to the flux taken into account. 
As mentioned in the main text, here we assume the gap in the superconductor is not affected by the normal-metal wire so that the coherent function takes its value in bulk superconductor,
\begin{equation}
  \gamma_{0}=-\frac{\Delta}{\epsilon+i\sqrt{\Delta^{2}-\epsilon^{2}}}.
\end{equation}

After solving the equations, we compute the spectral supercurrent 
\begin{equation}
j_E =\Re \left[ \frac {2(\tilde \gamma \partial \gamma - \gamma \partial \tilde \gamma)}{(1+\gamma \tilde \gamma)^2} \right],
\end{equation}
where $\tilde \gamma$ is the time reversed coherent function. Then we integrate the spectral supercurrent to find total supercurrent
\begin{equation}
I_s =\int j_E \tanh(\frac{\epsilon}{2T}) d\epsilon.
\end{equation}

Since the spectral current is a conserved quantity, this calculation can be done at any point of the wire. For wire segments 1 and 4, the total current is the supercurrent through the SNS SQUID, a dimensionless quantity normalized by $E_c /eR$,\cite{Wilhelm1997,Dubos2001a,Hammer2007} where $R$ is the normal resistance of wire of length $L$. The critical current at particular $\Phi$ is calculated by finding the maximum total current while  varying $\varphi$. With the leading order approximation, the critical current shows the conventional $|cos(\pi (\Phi/\Phi_0 ))|$ modulation. The energy of the system as a function of the flux $\Phi$ and phase difference $\varphi$ is then computed according to Eq.~(\ref{eqn_E_Phi}).
The results shown in Fig.~1, normalized by $(\hbar/2e^{2}R)E_{c}$, were computed for following parameters: $\Delta = 30$, $T =10$,  both normalized by $E_ c$, length is 1/3 for all 4 segments, normalized by $L$, and $A$ is [1,1,1,1] for the symmetric device and [1,1,0.75,1] for the asymmetric device, normalized by $R$ and cross-section area.

\end{document}